\documentclass[journal]{IEEEtran}
\usepackage{mathtools}

\usepackage{cite}

\ifCLASSINFOpdf
   \usepackage[pdftex]{graphicx}
   \DeclareGraphicsExtensions{.pdf,.jpeg,.png}
\else
   \usepackage[dvips]{graphicx}
   \graphicspath{{../eps/}}
   \DeclareGraphicsExtensions{.eps}
\fi
\usepackage{amsmath}
\usepackage{multirow}
\usepackage{array}

\usepackage{makecell}
\usepackage{xcolor}

\ifCLASSOPTIONcompsoc
 \usepackage[caption=false,font=normalsize,labelfont=sf,textfont=sf]{subfig}
\else
 \usepackage[caption=false,font=footnotesize]{subfig}
\fi

\usepackage{stfloats}

\begin{document}
%
\title{Real-Time Asphalt Pavement Layer Thickness Prediction Using Ground-Penetrating Radar Based on A Modified Extended Common Mid-Point (XCMP) Approach}
%
%
%

\author{Siqi~Wang, Zhen~Leng, Xin~Sui, Weiguang~Zhang, Tao~Ma and Zehui~Zhu
\thanks{This article has been accepted for publication in IEEE Transactions on Intelligent Transportation Systems. This is the author's version which has not been fully edited. Citation information: DOI 10.1109/TITS.2023.3343196}
\thanks{S. Wang, W. Zhang, and T. Ma are with the Department of Road Engineering, School of Transportation, Southeast University, Jiangning District, Nanjing, Jiangsu 211189 China.}
\thanks{Z. Leng and X. Sui are with the Department of Civil and Environmental Engineering, The Hong Kong Polytechnic University, Hung Hom, Kowloon, Hong Kong, SAR.}
\thanks{Z. Zhu is with the Department of Civil and Environmental Engineering, University of Illinois Urbana-Champaign, Urbana, IL, USA.}
}

%
%

\markboth{}%
{Shell \MakeLowercase{\textit{et al.}}: Bare Demo of IEEEtran.cls for IEEE Journals}
%



\maketitle

\begin{abstract}
The conventional surface reflection method has been widely used to measure the asphalt pavement layer dielectric constant using ground-penetrating radar (GPR). This method may be inaccurate for in-service pavement thickness estimation with dielectric constant variation through the depth, which could be addressed using the extended common mid-point method (XCMP) with air-coupled GPR antennas. However, the factors affecting the XCMP method on thickness prediction accuracy haven’t been studied. Manual acquisition of key factors is required, which hinders its real-time applications. This study investigates the affecting factors and develops a modified XCMP method to allow automatic thickness prediction of in-service asphalt pavement with non-uniform dielectric properties through depth. A sensitivity analysis was performed, necessitating the accurate estimation of time of flights (TOFs) from antenna pairs. A modified XCMP method based on edge detection was proposed to allow real-time TOFs estimation, then dielectric constant and thickness predictions. Field tests using a multi-channel GPR system were performed for validation. Both the surface reflection and XCMP setups were conducted. Results show that the modified XCMP method is recommended with a mean prediction error of 1.86\%, which is more accurate than the surface reflection method (5.73\%).
\end{abstract}


%
\IEEEpeerreviewmaketitle

\section{Introduction}
%
%
%
%
\IEEEPARstart{A}{ccurate} layer thickness prediction is required in the construction and maintenance stages of asphalt pavement \cite{mcdaniel2019impact, maser2006nde}. Ground-penetrating radar (GPR), an NDT technique that is based on the propagation and reflection of electromagnetic (EM) waves, has been successfully implemented to estimate asphalt pavement layer thickness \cite{de2018evaluation,wang2018continuous,lahouar2008automatic}. The single-channel antenna has been used for line-scan surveys to obtain the thickness profile on the survey track \cite{wang2022automatic,le2007thin,frid2022features}. The GPR antenna arrays allow multiple channels to perform surveys with lane-width coverage, which is more efficient than the pulse antenna \cite{liu2014situ,liang2022automatic,hartikainen2018algorithm,liu2018quantitative}. In addition, the antenna arrays enable transmitting and receiving EM waves from various channels, allowing different GPR data collection setups in addition to the conventional surface reflection method.

Thickness prediction using GPR is based on the EM wave travel time, or time-of-flight (TOF) within the asphalt layer, governed by the layer’s dielectric constant. This can be obtained based on the transmission line method using in-situ cores \cite{daniels2004ground}. Despite high accuracy, this method can only cover limited areas. The surface reflection method is usually used to estimate in-situ pavement dielectric constant. It requires the pavement surface reflection amplitude in the GPR signal and the calibration signal amplitude obtained using the EM reflection from a metal plate on the pavement surface \cite{spagnolini1997permittivity,astm}. This method has been widely implemented due to its simplicity; only one antenna with a transmitter and receiver is required \cite{varela2014semi,sun2021time,wang2020factors}. However, it assumes the uniformity of layer dielectric property through depth, which may cause inaccurate dielectric constant and thickness estimations for in-service pavement \cite{wang2020real,solla2021review,lai2018review,benedetto2017overview}. 

The common mid-point method (CMP) uses at least two GPR transmitter-receiver pairs to calculate the bulk dielectric constant based on the average propagation speed of the EM wave within the asphalt layer. Hence, it is independent of the dielectric property variation through the depth. The CMP method was first developed using two ground-coupled antennas, then modified using an air-coupled and ground-coupled antenna \cite{lahouar2002approach}. Despite high accuracy in predicting the bulk dielectric constant, the survey speed was limited due to the use of the ground-couple antenna. This is addressed by using two air-coupled antennas based on the extended CMP, or XCMP method, developed by Leng. The XCMP method could allow traffic-speed and non-contact pavement survey \cite{leng2014innovative}. Inputs such as the geometric information and EM wave TOFs within the asphalt layer are required to implement this method \cite{liu2023asphalt,warren2016gprmax}. 

Although it effectively estimates the bulk dielectric constant of the asphalt layer, the current XCMP method requires further investigations and improvements to allow real-time thickness predictions in field applications. The geometric information of the transmitter-receiver pairs and the TOFs of antenna channels need to be estimated with high accuracy to ensure reliable dielectric constant and thickness measurements. However, the effects of these input factors haven't been thoroughly studied for the XCMP method. The geometric information could be obtained during the calibration process before field testing. However, the TOFs need to be estimated during field tests to allow real-time results delivery. Currently, the TOFs are measured manually. An automatic algorithm is required to allow real-time dielectric constant and thickness predictions during GPR data collection using the XCMP method. The time-gating method recorded in \cite{wang2018continuous} may be used. However, it requires design thickness as known information, which may change during the in-service period.

This study investigates the affecting factors and develops a modified XCMP method to allow real-time in-situ asphalt pavement layer thickness prediction. A sensitivity analysis was performed to study the effects of layer dielectric constant variation, geometric information, and TOFs on the thickness prediction accuracy using the XCMP method, which is accomplished based on the finite-difference time-domain (FDTD) simulations using gprMax \cite{warren2016gprmax}. Dielectric constant and thickness were calculated using the surface reflection and XCMP methods. Results from both methods were compared based on the sensitivity analysis. A modified XCMP method was proposed using the edge detector to locate the asphalt layer surface and bottom reflection signals for TOF estimations. This could allow real-time dielectric constant and thickness estimations. The edge detection accuracy (EDA) was evaluated to identify an optimal sensitivity threshold for the automatic execution of the edge detector. Field tests using a 3D GPR system were performed. Results from conventional surface reflection and modified XCMP methods were analyzed for validation, and in-situ implementation suggestions were provided. 

The remainder of this paper is organized as follows: Section II briefly reviews the principles of the conventional surface reflection and XCMP methods for asphalt layer thickness measurement; Section III introduces the FDTD simulation setup; Section IV discusses the factors that affect thickness prediction accuracy using the XCMP method; Section V proposes an automatic TOF estimation method for automatic execution of the XCMP method in the field; Section VI provides field validation of the proposed method; and Section VII summarizes the conclusions and discusses future research directions.

\section{Thickness Prediction Principles Using GPR}
The prediction of pavement thickness is based on the TOF of EM waves within the layer, which is a function of the layer dielectric constant, as shown in Equations \ref{eqn_1} and \ref{eqn_2}:

\begin{equation}
        v=\frac{c}{\sqrt{\varepsilon}}
    \label{eqn_1}
\end{equation}

\begin{equation}
        D=\frac{vt}{2}
    \label{eqn_2}
\end{equation}

where $D$ is the layer thickness; $t$ is the TOF of EM wave within the layer, which can be estimated by the time delay between the layer surface and bottom reflections in the GPR signal; $v$ is the propagation speed of EM wave in $m/s$; $c$ is the speed of EM wave in free space, which is 0.3 $m/ns$; and $\varepsilon$ is the layer dielectric constant. The surface reflection method has been widely implemented to estimate asphalt layer dielectric constant, as shown in Equation \ref{eqn_3} \cite{lahouar2008automatic}:

\begin{equation}
        \varepsilon_1=(\frac{1+\frac{A_0}{A_{inc}}}{1-\frac{A_0}{A_{inc}}})^2
    \label{eqn_3}
\end{equation}

where $A_0$ is the reflection amplitude from the pavement layer surface; $A_{inc}$ is incident signal amplitude, which can be obtained based on the reflection from a flat metal plate (a perfect electrical conductor) on the pavement surface ($A_0/A_{inc}$  is referred to as the reflection coefficient); and $\varepsilon_1$ is the dielectric constant of the AC layer. However, Equation \ref{eqn_3} may not be accurate when applied to in-service asphalt pavement. Based on Snell's law, the estimated dielectric constant is the bulk value within the skin depth of penetration. The skin depth is the penetration distance where the amplitude of the EM wave reduces to $1/e$ ($e$ is the natural logarithm), or approximately 37\% of the original amplitude, as shown in Equation \ref{eqn_4} \cite{annan2005ground}:

\begin{equation}
        \delta=\frac{1}{\alpha}
    \label{eqn_4}
\end{equation}

where $\delta$ is the skin depth in $m$; and $\alpha$ is the attenuation factor in $1/m$. Skin depth is a function of the media conductivity and frequency. The penetration depth decreases with the increase in conductivity and frequency. The skin depth is usually less than 5 cm for the asphalt pavement material. Hence, the surface reflection method may be inappropriate when the pavement layer is thicker than 5 cm or in-service pavement with dielectric constant variation along the depth. 

\begin{figure}[!t]
    \centering
    \includegraphics[width=3.4in]{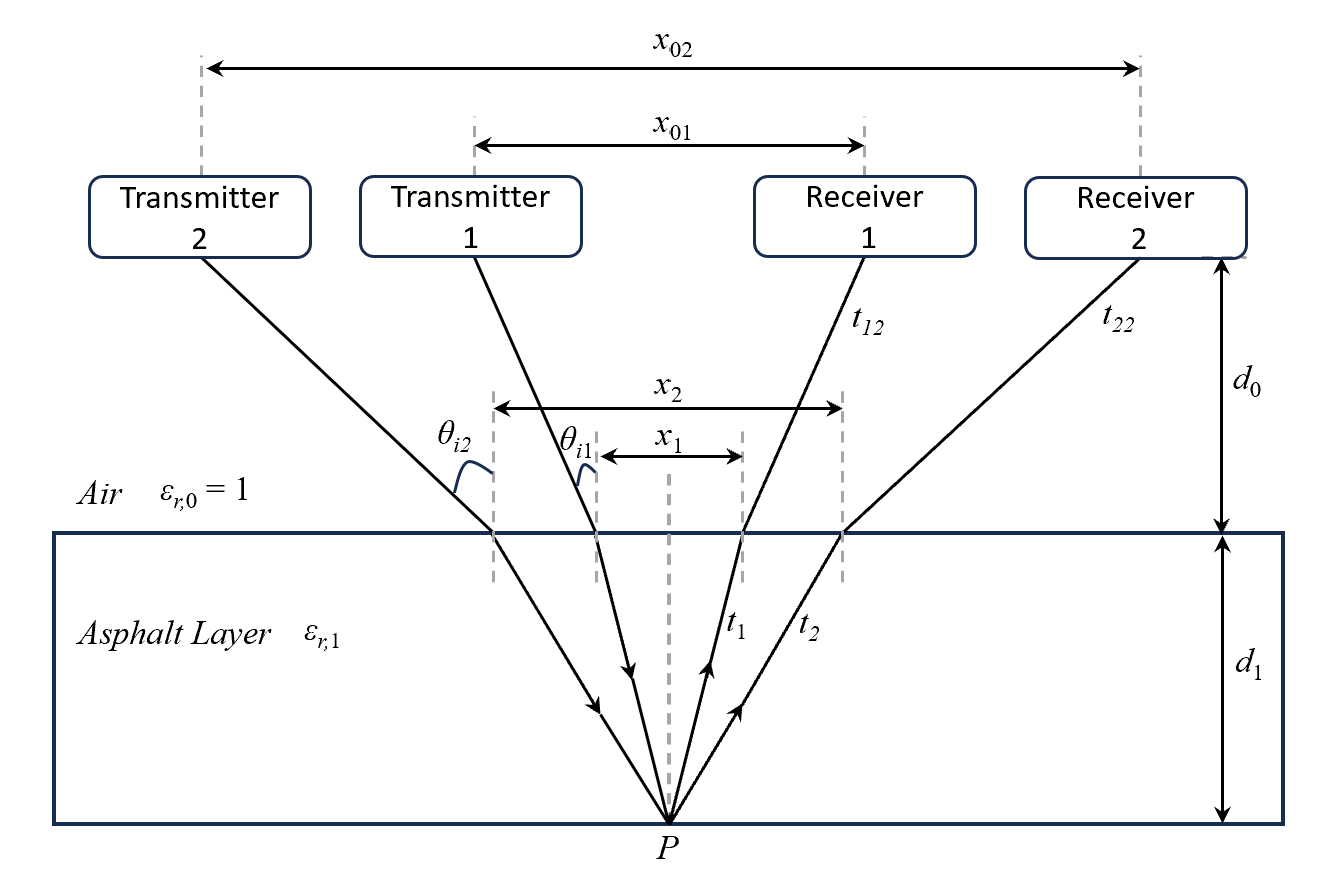}
    \caption{Diagram of EM wave propagation path using the XCMP method [24].}
    \label{fig_1}
\end{figure}

The XCMP method uses at least two air-coupled transmitters and receiver pairs, which can be reverted in one antenna. Figure \ref{fig_1} shows the EM wave propagation path in the XCMP setup \cite{leng2014innovative}. Waves are reflected from both the asphalt layer surface and bottom interfaces. Geometric information is required, including separation distances of the inner ($x_{01}$) and outer ($x_{02}$) antenna pairs, and the height from the antenna geometric center to the ground surface ($d_0$). Other information is required, such as the distance between the incident point and the reflection point of both antenna pairs ($x_1$ and $x_2$), incident angles $\theta_{i1}$ and $\theta_{i2}$, dielectric constant, and thickness of the asphalt layer. After mathematical deductions, the following equations can be obtained:

\begin{equation}
        t_1=\Delta t_1+\frac{\sqrt{4d_0^2+x_{01}^2}}{c}-\frac{\sqrt{4d_1^2+(x_{01}-x_1)^2}}{c}
    \label{eqn_5}
\end{equation}

\begin{equation}
        t_2=\Delta t_2+\frac{\sqrt{4d_0^2+x_{02}^2}}{c}-\frac{\sqrt{4d_1^2+(x_{02}-x_2)^2}}{c}
    \label{eqn_6}
\end{equation}

\begin{equation}
        (\frac{x_{01}-x_1}{2d_0})^2+1=\frac{t_1^2(x_2^2-x_1^2)^2}{t_1^2(x_2^2-x_1^2)^2-x_1^2c^2(t_2^2-t_1^2)^2}
    \label{eqn_7}
\end{equation}

\begin{equation}
        (\frac{x_{02}-x_2}{2d_0})^2+1=\frac{t_2^2(x_2^2-x_1^2)^2}{t_2^2(x_2^2-x_1^2)^2-x_2^2c^2(t_2^2-t_1^2)^2}
    \label{eqn_8}
\end{equation}

\begin{equation}
        \varepsilon_{AC}=\frac{c^2(t_2^2-t_1^2)}{x_2^2-x_1^2}
    \label{eqn_9}
\end{equation}

\begin{equation}
        d_1=\sqrt{(\frac{ct_1}{2\sqrt{\varepsilon_{AC}}})^2-(\frac{x_1}{2})^2}
    \label{eqn_10}
\end{equation}

where $\Delta t_1$ and $\Delta t_2$ are the TOF of EM waves obtained from surface and bottom reflections in the GPR signals; $t_1$ and $t_2$ are the TOF of EM waves from two antenna pairs within the asphalt layer. To use these equations, $\Delta t_1$ and $\Delta t_2$ should be obtained from GPR signals, then plugged into Equations \ref{eqn_5} and \ref{eqn_6} to obtain $\Delta t_1$ and $\Delta t_2$. Equations \ref{eqn_7} and \ref{eqn_8} can be numerically solved to obtain the distances between the incident and reflection points of two antenna pairs, which are $x_1$ and $x_2$. Then the dielectric constant and thickness of the asphalt layer can be calculated using Equations \ref{eqn_9} and \ref{eqn_10}.

Equations \ref{eqn_5}-\ref{eqn_10} indicate that affecting factors of the XCMP method include the antenna geometric setups, mounting positions, and the TOF from the surface and bottom reflections in the GPR signals. The effects of these factors on dielectric constant and thickness prediction accuracies using the XCMP method were explored via numerical simulations in this study.

\section{Simulation Model}
The numerical simulation model was built using an open-source tool called the gprMax, which has been implemented to study the EM wave reflection features in the GPR applications \cite{warren2016gprmax}. It solves the Maxwell equations in the time domain using the finite-difference time-domain (FDTD) method \cite{yee1966numerical}. This study analyzed the effects of layer dielectric constant variation, antenna geometric setup, and TOFs on the thickness prediction accuracy using the XCMP method by comparing the dielectric constant and thickness estimations to those using the surface reflection method and ground-truth values in the model. This eliminates the effect of external noise in the field test.

\subsection{Geometric and Electrical Properties}
The model was built in two dimensions \cite{wang2020real,wang2020factors}, which could significantly reduce the computational expense compared to three-dimensional numerical models. This can meet the simulation accuracy because the locations and amplitudes of interface reflections are the same between two-dimensional and three-dimensional models; the area of impact of the air-coupled antenna on the pavement surface can be assumed to be identical. The model was a rectangle with 2 $m$ in width and 1.2 $m$ in height. Three layers, namely, air (100 $cm$), asphalt pavement (10 $cm$), and the base layer (10 $cm$) were built in the model. These geometric parameters were chosen to simulate the field test scenarios based on previous research \cite{leng2014innovative}. 

The gprMax tool allows the setting of each layer’s electrical properties. The electric permeabilities and conductivities of the asphalt and base layers were defined as one and zero, which assumes the non-magnetic, non-dispersive, and non-conductive properties of the asphalt pavement construction materials, suggested by previous researchers \cite{plati2020integration}. The dielectric constant of the base layer was 10. The dielectric constant variation of the asphalt layer was simulated using five sublayers through the depth. Both the increasing and decreasing dielectric constant distributions were simulated, as shown in Table  \ref{table_1}. The range of both increasing/decreasing  gradients was defined based on in-situ asphalt layer dielectric constant measurements\cite{wang2020factors}.

\begin{table}[!t]
\centering
\renewcommand{\arraystretch}{1.3}
\caption{Dielectric Constant Values of Sublayers in the Simulated Pavement Structure.}
\label{table_1}
\begin{tabular}{|c|c|c|}
\hline
Dielectric Constant Gradients     & Decreasing Trends & Increasing Trends \\ \hline
Sublayer 1& 6          & 5.2        \\ Sublayer 2& 5.8        & 5.4        \\ Sublayer 3& 5.6        & 5.6        \\ Sublayer 4& 5.4        & 5.8        \\ Sublayer 5& 5.2        & 6          \\ \hline
\end{tabular}
\end{table}

\subsection{Antennas}
The antenna central frequency affects the signal penetration depth and resolutions \cite{annan2005ground}. In this simulation, a pulse antenna with a central frequency of 2 GHz is considered. This differs from the stepped frequency antenna used in the field validation in this study. This central frequency was selected because of its high vertical resolution and sufficient penetration depth\cite{Wang2023DetectabilityOC}. The pulse antenna could provide a simpler waveform than the stepped frequency antenna in the time domain. The main and side gradients in the pulse antenna waveform can be located using a one-dimensional edge detector. On the contrary, the waveform from a stepped frequency antenna contains different frequency components in the time domain, which has ripples that could decrease the gradient evidence, leading to inaccurate localization using edge detectors. Hence, simulating the pulse antenna waveform is beneficial for capturing signal gradients in the time domain using the proposed automatic method, which is validated using stepped frequency signals in the field test in a later context.

\begin{figure}[!t]
    \centering \includegraphics[width=3.4in]{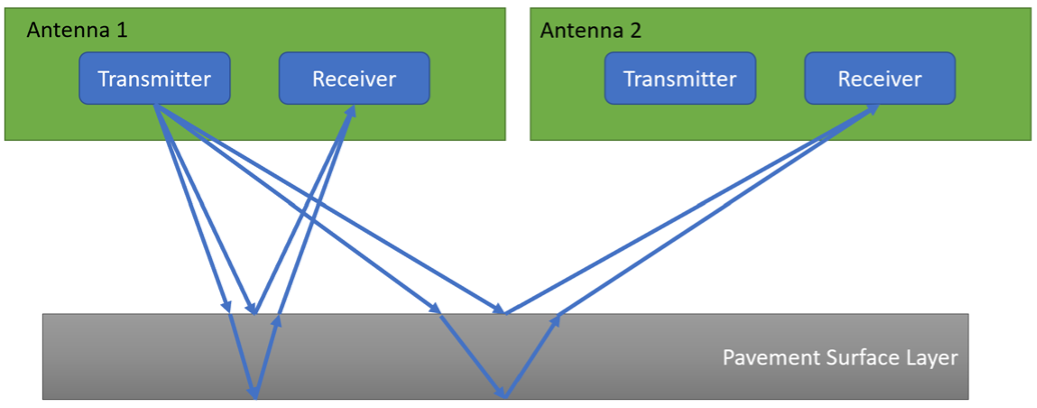}
    \caption{EM wave paths of the simulated XCMP setting.}
    \label{fig_2}
\end{figure}

The numerical model simulated the XCMP and surface reflection methods as signal collection modes using air-coupled antennas. Both the transmitters and receiver pairs were simplified as dipole antennas. They were positioned 80 $cm$ above the asphalt layer surface. The outer and inner sets of dipole offsets were 120 $cm$ and 40 $cm$, respectively. Figure \ref{fig_2} shows the expected EM wave paths, which differs from Figure \ref{fig_1}. This antenna setting aims to simulate the receiving of EM wave by two receivers from one transmitter, which ensures sufficient TOF differences from both antenna pairs ($\Delta t_1$ and $\Delta t_2$ in Equations \ref{eqn_5} and \ref{eqn_6}), suggested by Leng in \cite{leng2014innovative}. Meanwhile, the transmitted and received signals from the inner antenna pair were used to calculate the dielectric constant and thickness using the surface reflection method. This is valid because 40 cm is close to the offset between the transmitter and receiver of a horn antenna when the surface reflection method is applied \cite{annan2005ground}.

\subsection{Signal Decoupling}
\begin{figure*}[!t]
\centering
\subfloat[Before decoupling]{\includegraphics[width=3.3in]{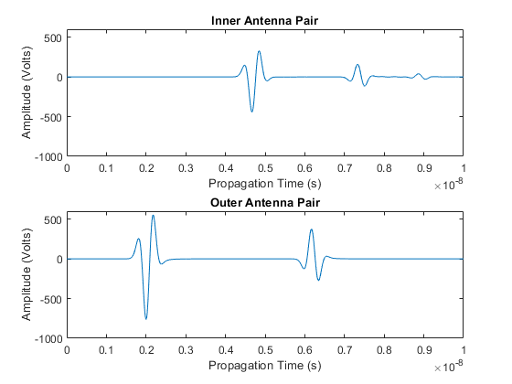}}
\label{fig_3a}
\hfil
\subfloat[After decoupling]{\includegraphics[width=3.3in]{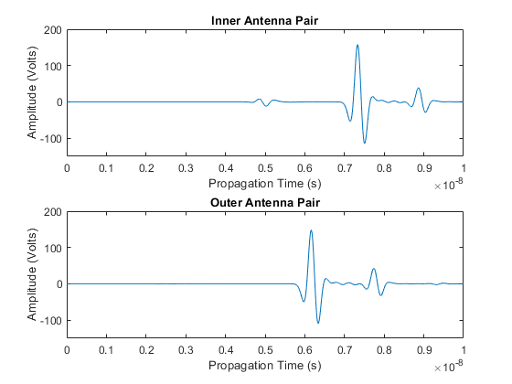}
\label{fig_3b}}
\caption{A-Scan signals from the inner and outer antenna pairs.}
\label{fig_3}
\end{figure*}

Due to the simplification of the transmitter and receiver antennas as dipoles, a coupling pulse is observed with a significantly larger amplitude than pavement layer interface reflections. This is due to the lack of shielding between the transmitter and receiver dipoles, which allows an intensive direct EM energy transmission between the two, as shown in Figure \ref{fig_3}\emph{(a)}. This was addressed by building an additional model using air in the entire numerical model domain. The signal received from this model is the coupling signal, which simulates shooting the antenna toward the sky during field data collection. Then the decoupled signal was obtained by subtracting the coupling signal from the signals in Figure \ref{fig_3}\emph{(a)}, as shown in Figure \ref{fig_3b}\emph{(b)}. The time delays, or TOFs between reflections, are EM wave propagations in the asphalt layer from the inner and outer antenna pairs. These TOFs and geometric information are inputs to calculate the dielectric constant and thickness using the XCMP method.

\section{Factors Affecting Thickness Prediction Accuracy Using the XCMP Method }
The mathematical equations of the XCMP method indicate several factors that may affect the dielectric constant and thickness prediction accuracies. These factors include the asphalt layer dielectric depth variation, antenna geometric setup, and estimated TOFs from the asphalt layer surface and bottom sides in the GPR signals from both antenna pairs. The effects of these factors are discussed as follows.

\subsection{Asphalt Layer Dielectric Constant Depth Variation}
The thickness prediction accuracies were compared between the XCMP and surface reflection methods via a layer dielectric constant model with variation through the depth. The asphalt layer consists of five sub-layers that simulate the increasing and decreasing dielectric constant variations ranging from 5.2 to 6. Table \ref{table_2} summarizes the asphalt layer's calculated dielectric constant and thickness results using the surface reflection and XCMP methods. The predicted dielectric constant using the XCMP method is 5.6, which equals the model's mean value of the asphalt layer. However, dielectric constant prediction using the surface reflection method is higher than 5.6 in the decreasing variation case and lower than 5.6 in the increasing case. This inaccuracy comes from the principle of the surface reflection method, which only measures the averaged dielectric constant within the skin depth of the layer. As a result, the predicted thickness using the XCMP method (errors of 1.4\% and 1.5\% in the decreasing and increasing cases, respectively) is more accurate than that using the reflection method (errors of 6.4\% and 24.7\% in the decreasing and increasing cases, respectively). This shows that the XCMP method is more appropriate than the surface reflection method for thickness prediction of in-service pavement with dielectric property variation through depth.

\begin{table}[!t]
\centering
\renewcommand{\arraystretch}{1.3}
\caption{Dielectric constant and thickness predictions of the asphalt layer using the XCMP and reflection methods under increased/decreased dielectric constant gradients.}
\label{table_2}
\begin{tabular}{|c|c|c|}
\hline
  \makecell{Dielectric Constant and \\ Thickness Predictions} & Decreased Gradients&  Increased Gradients\\ \hline
  \makecell{Surface Reflection \\ Thickness (cm)} & 9.36      & 12.47                            \\ 
  XCMP   Thickness (cm)&10.14                    & 10.15                            \\ 
  \makecell{Surface Reflection \\ Dielectric Constant} & 6.41 & 5.30                             \\ 
  \makecell{XCMP \\ Dielectric Constant} & 5.60               & 5.60                             \\ \hline
\end{tabular}
\end{table}

\subsection{TOF Variation}
\begin{figure*}[!t]
\centering
\subfloat[Dielectric constant]{\includegraphics[width=3.3in]{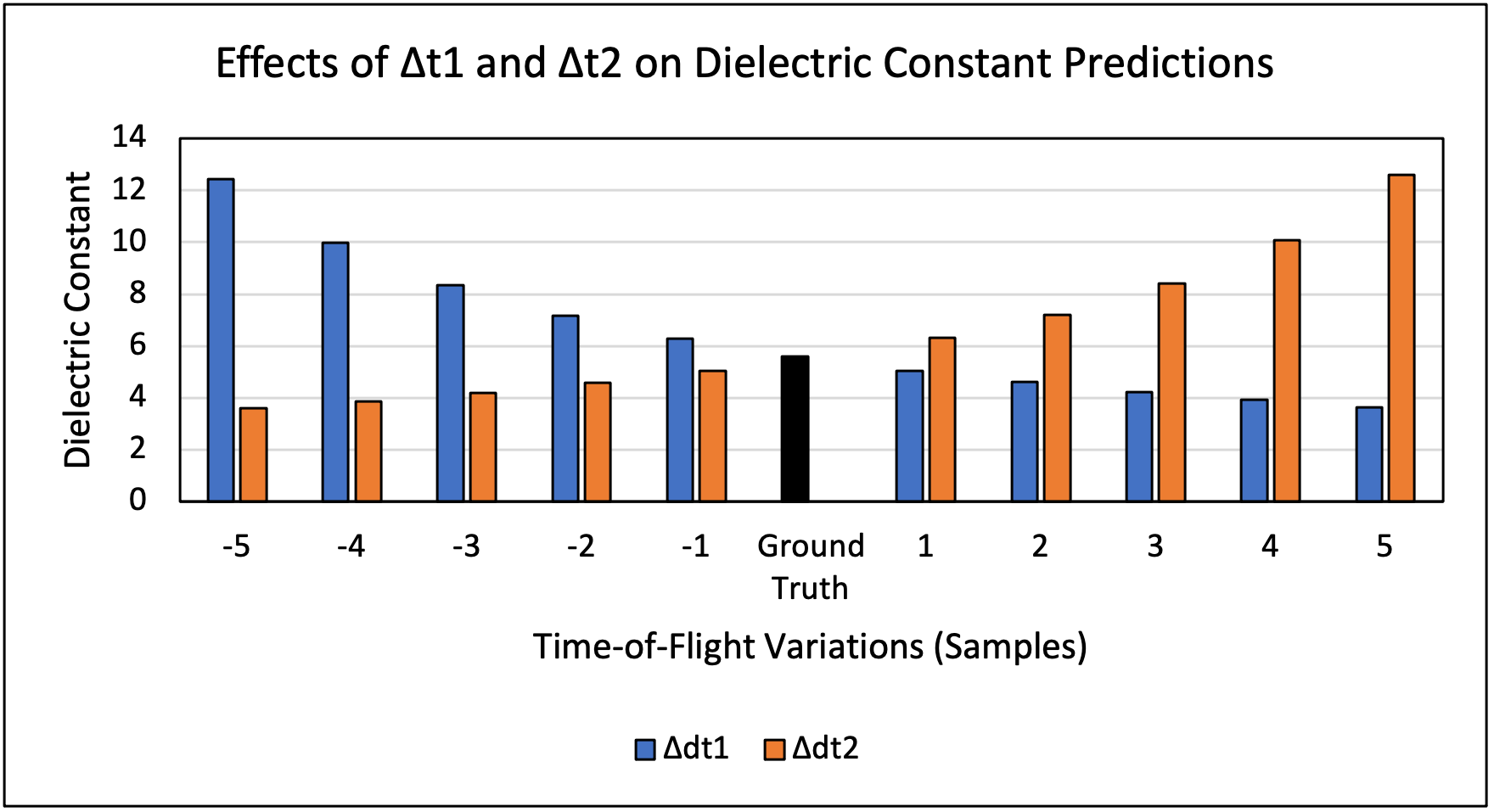}}
\label{fig_4a}
\hfil
\subfloat[Layer thickness]{\includegraphics[width=3.3in]{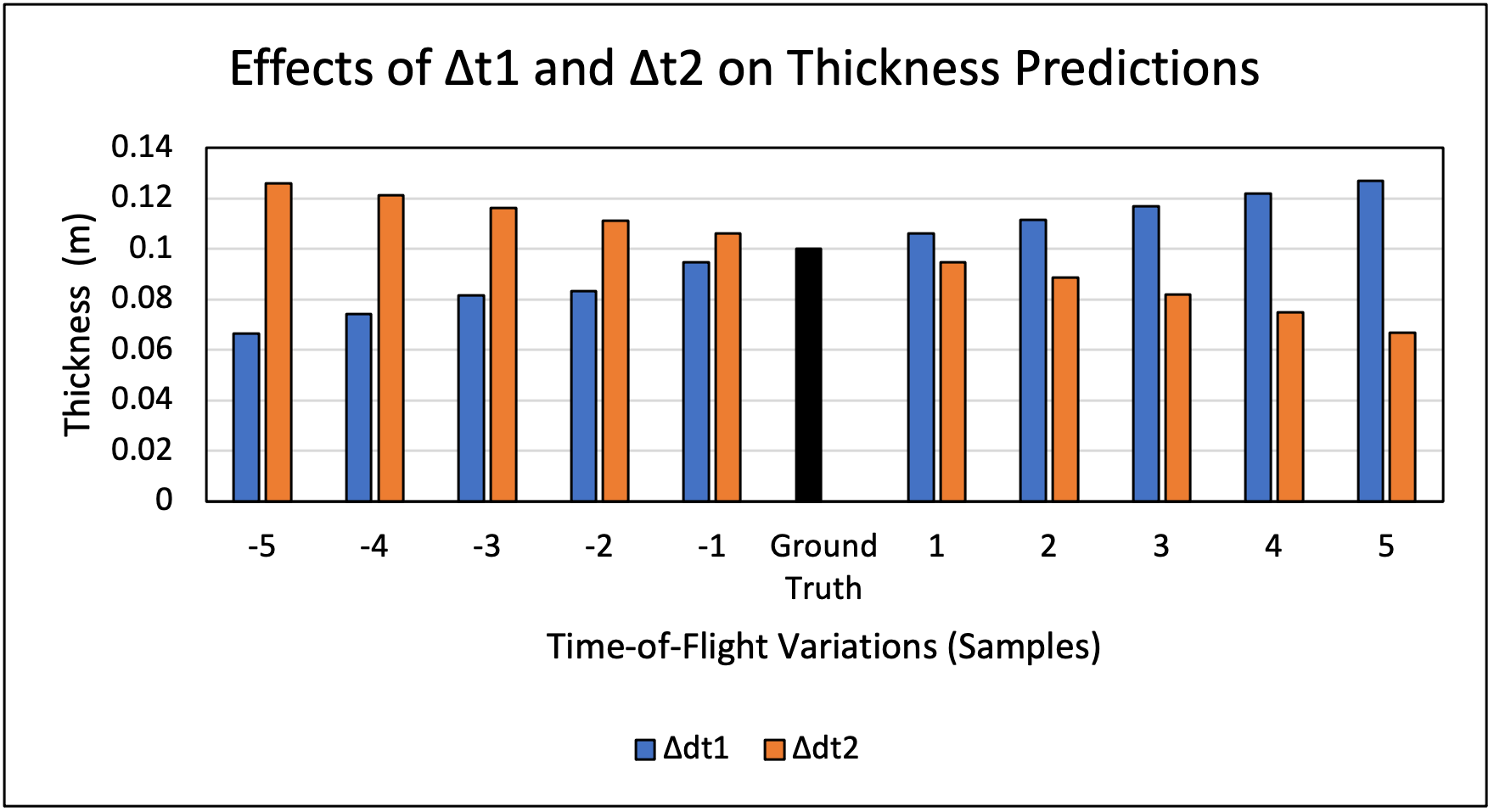}
\label{fig_4b}}
\caption{Effects of TOF variations on prediction results.}
\label{fig_4}
\end{figure*}

The TOFs between the reflected signals from the asphalt layer interfaces ($\Delta t_1$ and $\Delta t_2$) are required as inputs from both antenna pairs using the XCMP method. In the GPR signal, TOFs are identified by the samples' time stamps at the interface reflections’ peak. Hence, the TOF resolution, i.e., the minimum time index of the TOF that can be misidentified, is one sample divided by the time window of the A-Scan signal. For example, 2121 samples were generated in the simulated A-Scan signal with a time duration of 10 ns. Changing by one sample means altering the time interval by 10/2121 ns. The number of 2121 was based on the model's simulated time duration and step size. The FDTD simulation in gprMax executes the EM wave propagation in the time mode by central-difference spatial and temporal derivatives. Given the simulated time duration and minimum derivative step size in gprMax, the simulation steps can be obtained automatically, which is 2121.  Figure \ref{fig_4} summarizes the predicted dielectric constant and thickness results by changing $\Delta t_1$ and $\Delta t_2$ from one to five samples. The ground-truth values are shown in the figure for comparison. Results show that estimation errors by one sample of either $\Delta t_1$ or $\Delta t_2$ lead to a dielectric constant prediction error of 7\%, leading to a thickness estimation error of 10\% of the asphalt layer. This necessitates the accurate estimations of TOFs for both antenna pairs using the XCMP method.

\subsection{Antenna Pair Distance Variation}
\begin{figure*}[!t]
\centering
\subfloat[Dielectric constant]{\includegraphics[width=3.3in]{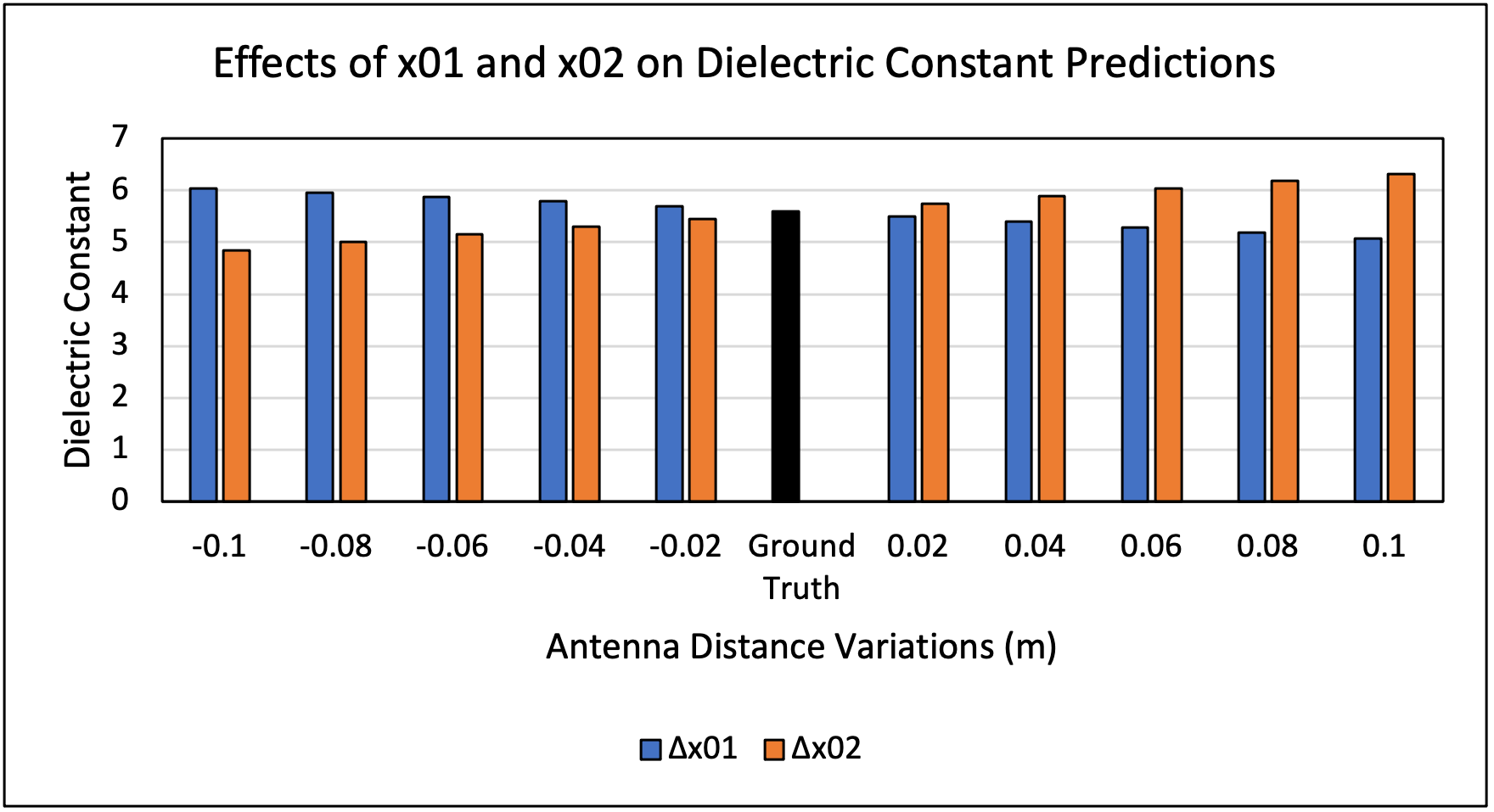}}
\label{fig_5a}
\hfil
\subfloat[Layer thickness]{\includegraphics[width=3.3in]{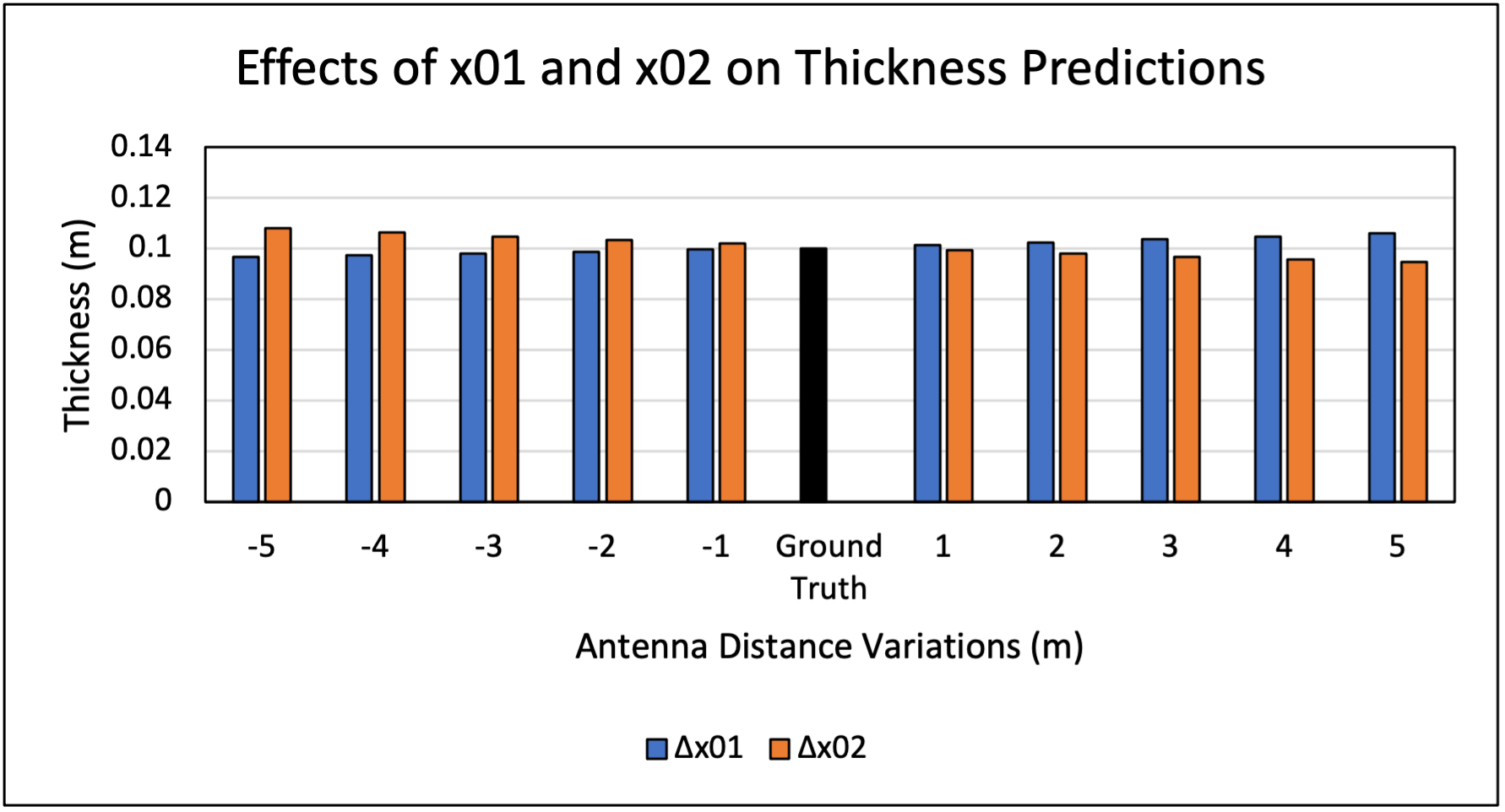}
\label{fig_5b}}
\caption{Effects of antenna distance variations on prediction results.}
\label{fig_5}
\end{figure*}

\begin{figure*}[!t]
\centering
\subfloat[Dielectric constant]{\includegraphics[width=3.3in]{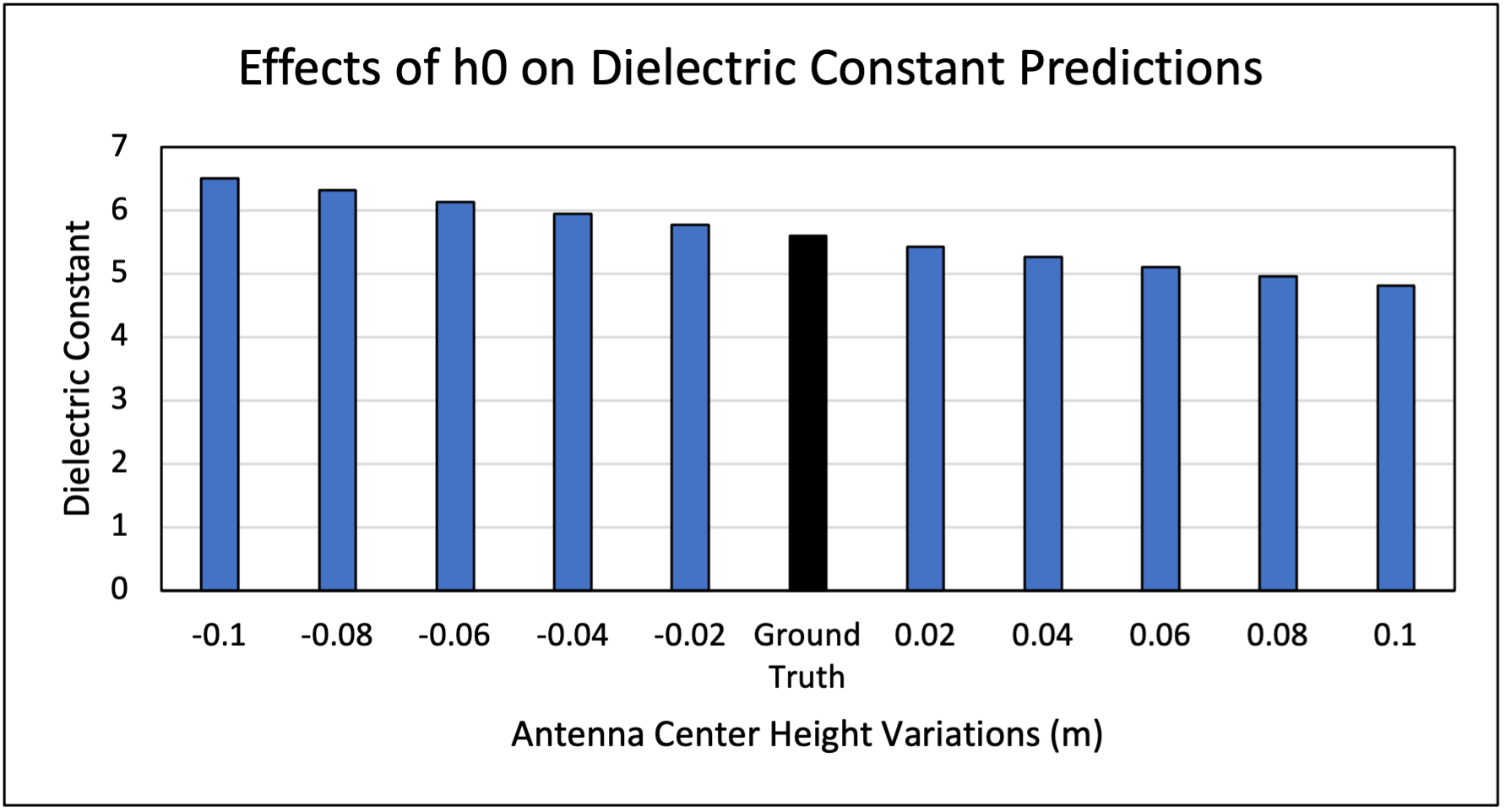}}
\label{fig_6a}
\hfil
\subfloat[Layer thickness]{\includegraphics[width=3.3in]{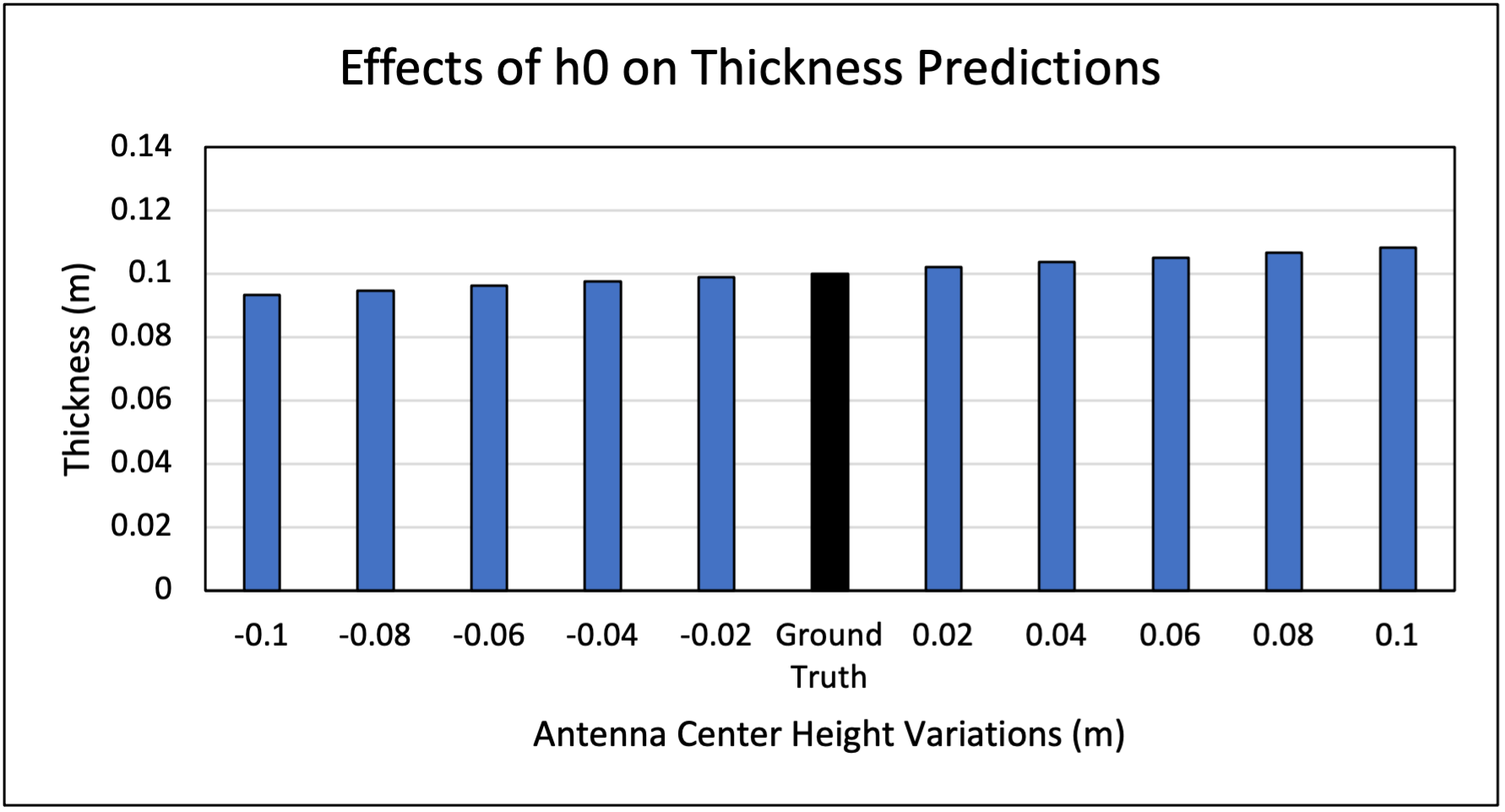}
\label{fig_6b}}
\caption{Effects of antenna center height variations on prediction results.}
\label{fig_6}
\end{figure*}

The geometric information required by the XCMP method includes the distance between the transmitter and receiver antennas of the outer ($x_{01}$) and inner ($x_{02}$) pairs, as well as the height of the antenna geometric center ($h_0$). The effects of the antenna pair distance on the dielectric constant and thickness predictions are shown in Figure \ref{fig_5}. An estimation error of either $x_{01}$ or $x_{02}$ by 2 $cm$ leads to a dielectric constant error of 2\% and a thickness error of 1\%, significantly less than that when changing one sample in the TOF of both antenna pairs. In addition, an estimation error of the antenna geometric center by 2 cm leads to a dielectric constant error of less than 2\% and a thickness error of less than 1\%, as shown in Figure \ref{fig_6}. Hence, the geometric information of antenna pairs has a negligible effect on the dielectric constant and thickness estimation accuracies compared to TOFs. 

In addition to TOF estimations and geometric information, the asphalt layer thickness and dielectric constant contrast between the asphalt and underlaid base layer may affect the thickness prediction accuracy using the XCMP method. Previous research has concluded that when the layer thickness is comparable to or less than the GPR incident signal wavelength, the reflection signals from both the asphalt layer surface and bottom sides may overlap with each other, contributing to dielectric constant and thickness prediction errors \cite{le2007thin,wang2020real,pan2018time}. For the dielectric property contrast, it was found that the reflection signal from the layer interface is evident when the dielectric constant difference is no less than one between two layers.

In summary, dielectric constant and thickness predictions using the XCMP method are more accurate than the conventional surface reflection method with dielectric property variation through the depth of the asphalt layer. For the XCMP method applications, the estimation accuracy of TOFs from both antenna pairs has a significantly higher effect than the geometric information (antenna offset and geometric center height) on the prediction accuracies. This necessitates the accurate estimation of TOFs. An automatic method is suggested to provide real-time signal processing when the XCMP method is implemented in the field, which is proposed in the next section to estimate the TOFs automatically.

\section{Automatic TOF Estimation}
The TOFs from both antenna pairs can be estimated by obtaining the time delay between the layer interface reflections, usually accomplished via manual inspection. An automatic and accurate TOF estimation approach is suggested in real-time field applications using the XCMP method. 

\begin{figure}[!t]
    \centering
    \includegraphics[width=3.4in]{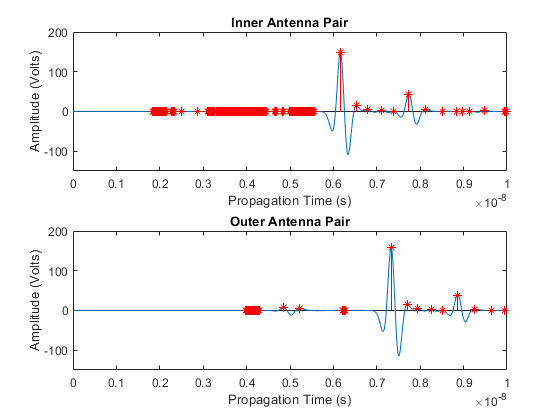}
    \caption{Maximum points from the A-Scan waveform.}
    \label{fig_7}
\end{figure}

The reflection signals from the asphalt layer surface and bottom interfaces are peaks in the GPR signals. They could be identified by locating the local maxima in the signals. Figure \ref{fig_7} shows the identified maxima points from both antenna pairs, highlighted in red. In addition to the peaks of layer surface and bottom reflections, points in small-scale ripple waveforms are also picked up, which may be misidentified as asphalt layer interface reflections. One possible reason is the numerical errors during the decoupling process by subtracting the coupling pulse from the raw GPR signal, for example, the highlighted points before 5.5 ns in the figure. The second parts of misidentified points are from dielectric property variation within the asphalt layer, for example, 6-7.8 ns from the inner pair and 7.3-8.8 ns from the outer pair. The time-gating method developed by \cite{wang2018continuous} may assist in locating the actual reflection signals from layer interfaces based on these identified points. However, this requires layer design thickness as prerequisite knowledge, which may change during the pavement in-service stage. 

Edge detectors may allow automatic identification of the asphalt layer interface reflections by capturing the signal gradient information. This outperforms the thresholding method because the absolute amplitude values of the GPR signal are voltages multiplied by a correction factor suggested by the vendors, which may vary among different GPR antenna models. Hence, the detected gradient information using edge detectors is independent of the correction factor, making it applicable to signals from different GPR models. Widely used kernel edge detectors include Sobel, Prewitt, Roberts, and Canny detectors. They have comparable performances with the default kernel size. The Sobel edge detector has been proven to be robust and efficient when applied to locating the reflection gradients in GPR signals for thickness prediction \cite{wang2022automatic}. Hence, only the Sobel edge detector was used in this study. The general form of this detector is shown in Equations \ref{eqn_11}-\ref{eqn_13}.

\begin{equation}
        L_x=\begin{bmatrix}
1 & 0 & -1\\
2 & 0 & -2\\
1 & 0 & -1
\end{bmatrix}
    \label{eqn_11}
\end{equation}

\begin{equation}
        L_x=\begin{bmatrix}
1 & 2 & 1\\
0 & 0 & 0\\
-1 & -2 & -1
\end{bmatrix}
    \label{eqn_12}
\end{equation}

\begin{equation}
        |\nabla L|=\sqrt{L_x^2+L_y^2}
    \label{eqn_13}
\end{equation}

where $L_x$ and $L_y$ are the gradient extractors at horizontal and vertical directions, respectively, used to extract edge features at corresponding directions; $|\nabla L|$ is the norm of the synthesized gradient. Edge features are obtained by convolving the kernel with the input signal, shown in Equation 
\ref{eqn_14}.

\begin{equation}
        k[m,n]=(f*g)[m,n]=\sum_{i,j}f[m-i,n-j]g[i,j]
    \label{eqn_14}
\end{equation}

where $k[m,n]$, $f[m,n]$ and $g[m,n]$ are the edge map, raw signal, and the kernel, respectively. Because the layer interface reflections have a significantly larger gradient than the rest of the waveforms in the GPR signals, they can be treated as edge features in the horizontal direction, which would be detected using the horizontal edge detector. The detected edge map can be displayed as markers in the GPR signals.

The sensitivity threshold is required as the input parameter for an edge detector in programming languages such as MATLAB and Python. The calculated amplitude gradients weaker than the sensitivity threshold should be removed. Hence, the sensitivity threshold value affects the detected edge features in the GPR signals. Usually, this value is decided heuristically, depending on the input signal. It can be identified by trial and error, which may impede the real-time signal processing and thickness prediction using GPR in the field.

An optimal sensitivity threshold specified for XCMP applications may be identified by comparing the detected edge marks and the actual locations of the asphalt layer surface and bottom reflections.  The edge detection accuracy (EDA) was calculated using the ratio between the lengths of reflection signal waveforms and the detected marks in the edge map. This value reaches one when the entire reflection waveforms fall into the range of detected edge marks, where the threshold reaches the optimal value. Normalization of the input GPR signal is suggested to execute the edge detector. Because the threshold is determined based on the amplitude gradient of the GPR signal, normalization makes the identified threshold independent of the GPR signal gain, which determines the absolute values of the reflection gradients. The normalization was executed by dividing the A-Scan signal by the maximum reflection amplitude.

\begin{figure*}[!t]
\centering
\subfloat[0.005]{\includegraphics[width=3.3in]{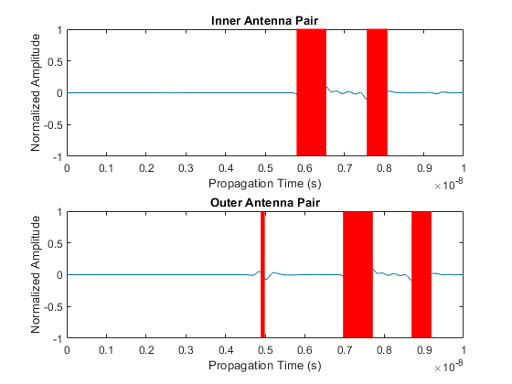}}
\label{fig_8a}
\hfil
\subfloat[0.01]{\includegraphics[width=3.3in]{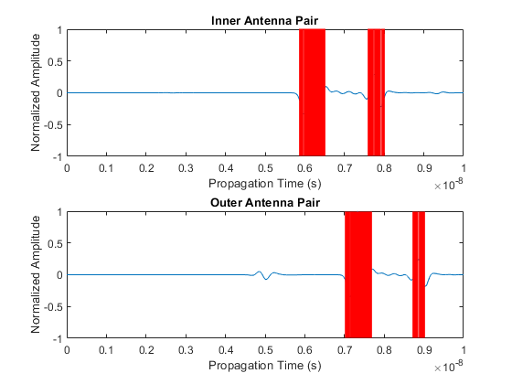}
\label{fig_8b}}
\hfil
\subfloat[0.02]{\includegraphics[width=3.3in]{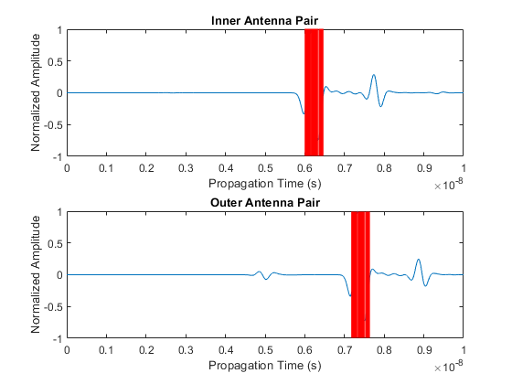}
\label{fig_8c}}
\caption{Edge detection results using thresholds of 0.005, 0.01, and 0.02.}
\label{fig_8}
\end{figure*}

\begin{figure}[!t]
    \centering
    \includegraphics[width=3.4in]{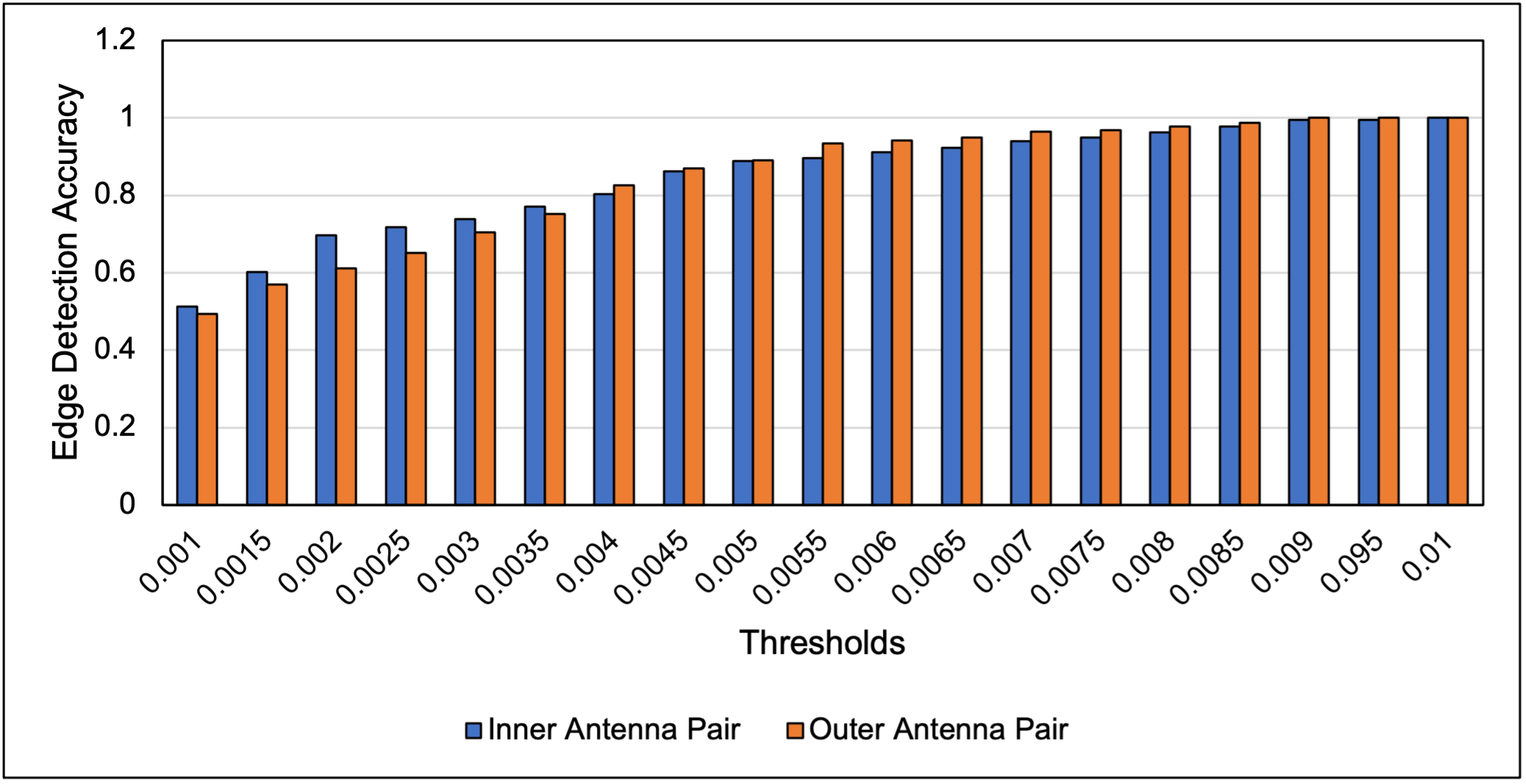}
    \caption{Edge detection accuracy results with respect to threshold increase using the Sobel edge detector.}
    \label{fig_9}
\end{figure}

\begin{figure}[!t]
    \centering
    \includegraphics[width=3.4in]{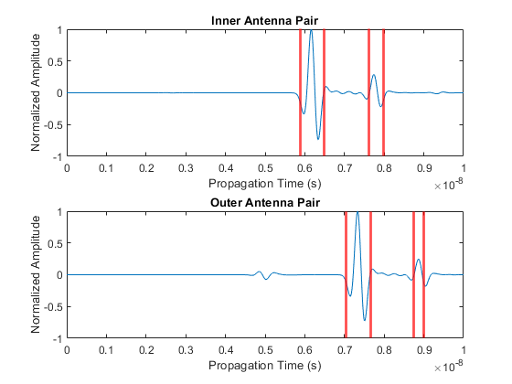}
    \caption{Detected layer interface reflections using the edge detector.}
    \label{fig_10}
\end{figure}

Figure \ref{fig_8} shows examples of detected edge maps overlaid on the signals with sensitivity thresholds of 0.005, 0.01, and 0.02. As the threshold increases, the detected edge features accumulate on the asphalt layer surface and bottom reflections. This is because these two reflections in the signal are the most significant ones with the highest amplitude gradients. As the threshold reaches 0.01 in Figure \ref{fig_8}, the EDA value reaches one for both antenna pairs, summarized in Figure \ref{fig_9}. The detected features start to miss the layer bottom reflection as the threshold goes beyond 0.01, for example, 0.02 in Figure \ref{fig_8}\emph{(c)}. Increasing the threshold monotonically decreases the accuracy of detected reflection gradients in the signal. Hence, an optimal threshold is identified as 0.01. This value can be applied to signals from both antenna pairs because it is independent of the absolute value of the signal amplitudes.
Figure \ref{fig_8}\emph{(b)} shows the detected feature maps accumulated around the reflection waveforms from the asphalt layer surface and bottom sides. Ripples before the layer surface reflection and within the asphalt layer are ignored, outperforming the detected points using local maxima in Figure \ref{fig_7}. Only the start and end markers were picked at two edge feature clusters in each signal to demonstrate the detected reflection signals further, as shown in Figure \ref{fig_10}. Then the mean value of the time stamps of the start and end markers was calculated as the reflection location. The TOFs are calculated based on the time delay between the surface and bottom signal locations.

In summary, the XCMP method may be used to estimate the dielectric constant and thickness of the asphalt layer in field applications, especially for in-service pavement sections with dielectric property variation through depth, which outperforms the conventional surface reflection methods. At least two air-coupled antennas or a multi-channel antenna array with the same central frequencies are required. The following pipeline is suggested:
\begin{itemize}
    \item Obtain the geometric information, e.g., antenna center height ($d_1$) and offset between transmitters and receivers of two antenna pairs ($x_{01}$ and $x_{02}$).
    \item Collect GPR signals. Perform the proposed method to obtain the TOFs from both the antenna pairs as $\Delta t_1$ and $\Delta t_2$.
    \item Obtain EM wave propagation time of both antenna pairs, $t_1$ and $t_2$.
    \item Numerically solve Equations \ref{eqn_7} and \ref{eqn_8} to obtain the distance between the incident point and the reflection point of both antenna pairs ($x_1$ and $x_2$).
    \item Predict the bulk dielectric constant and thickness of the asphalt layer.
\end{itemize}

\section{Field Validations}
\begin{figure*}[!t]
\centering
\subfloat[Section 1]{\includegraphics[width=2.3in]{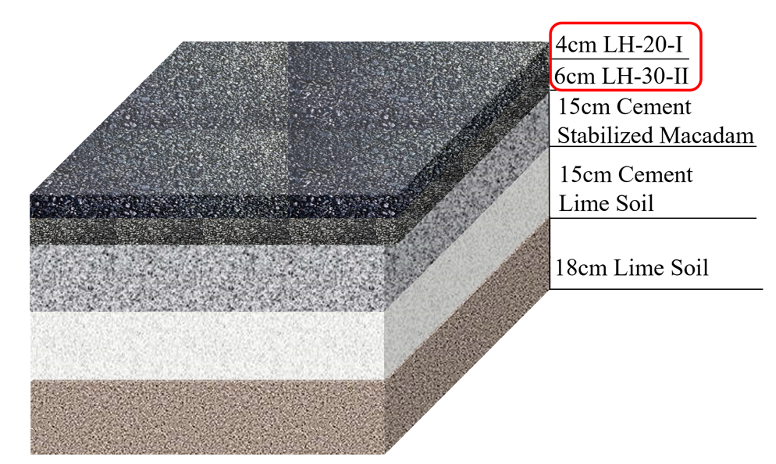}}
\label{fig_11a}
\hfil
\subfloat[Section 2]{\includegraphics[width=2.3in]{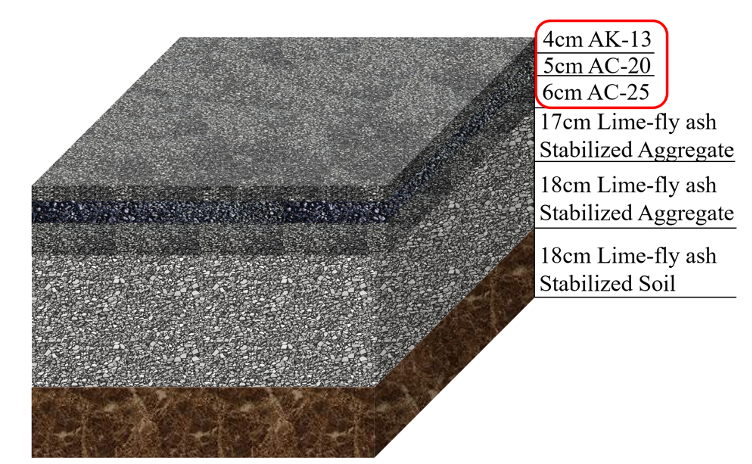}
\label{fig_11b}}
\hfil
\subfloat[Section 3]{\includegraphics[width=2.3in]{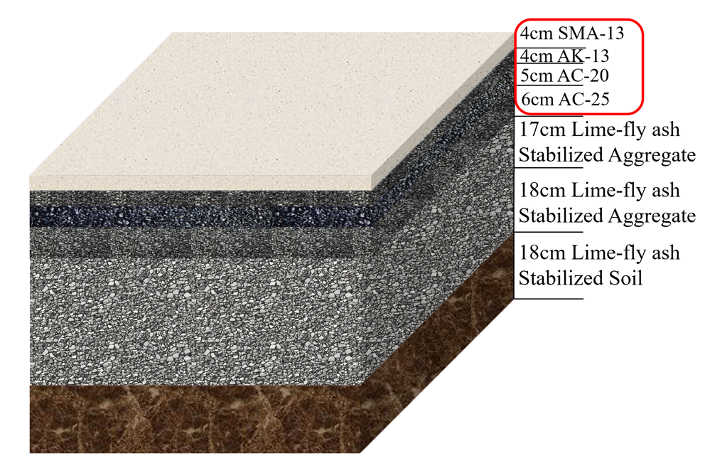}
\label{fig_11c}}
\caption{Pavement cross-sections of the surveyed sections.}
\label{fig_11}
\end{figure*}

The proposed modified XCMP method was validated using field test data in this study, conducted on Hebao Highway in Shandong and Henan Province, China. The objective was to investigate layer thickness distributions after 20 years of service. The design pavement structures of the surveyed sections are shown in Figure \ref{fig_11}. As highlighted in the figure, the design thicknesses of the asphalt layer were 10 $cm$, 15 $cm$, and 19 $cm$, respectively. However, the thickness may change during the in-service period. This is beneficial for validating the proposed algorithm because obtaining the thickness is a 'blind test' using the proposed approach without known information of actual thickness. The proposed edge detection method in the simulation study can be implemented to obtain the TOFs from each transmitter-receiver pair automatically. Due to the negligible dielectric constant contrasts between each sub-layer in the surface layer component, two or three sublayers were combined for GPR data interpretation. 

\begin{figure}[!t]
    \centering
    \includegraphics[width=3.4in]{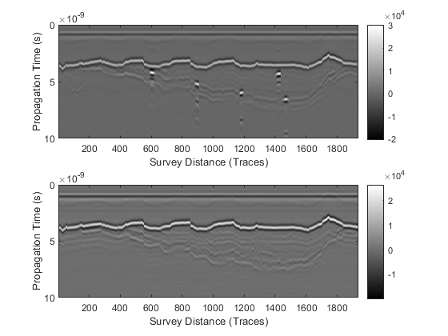}
    \caption{B-Scan images of GPR raw data from two transmitter-receiver pairs using the modified XCMP method.}
    \label{fig_13}
\end{figure}

Data was collected using a multi-channel stepped frequency GPR system. The antenna was a DXG1212 model with a frequency bandwidth from 200 MHz to 3 GHz. The sampling interval was 7.5 cm between two consecutive signal traces. This could cover the diameter of in-situ cores, which was used to validate the thickness prediction accuracy using GPR. Data is collected using surface reflection and XCMP methods to compare the thickness prediction accuracies. During data collection, the sampling rates of the surface reflection and XCMP methods are 512 and 1024 samples, respectively. The objective of using 1024 in the XCMP method is to achieve a higher TOF resolution than the rate of 512 used in the surface reflection method. The pavement sections comprised a passing lane, a driving lane, and a road shoulder. Due to traffic control limitations, GPR surveys were conducted on the driving lane with a coverage width of 0.9 m. Raw GPR data was B-Scan images from each antenna pair. The location of transmitters and receivers was recorded and mapped to in-situ locations using onboard odometers. B-Scan images on the survey tracks are shown in Figure \ref{fig_13}.

\begin{figure*}[!t]
\centering
\subfloat[GPR survey tracks]{\includegraphics[width=2.5in]{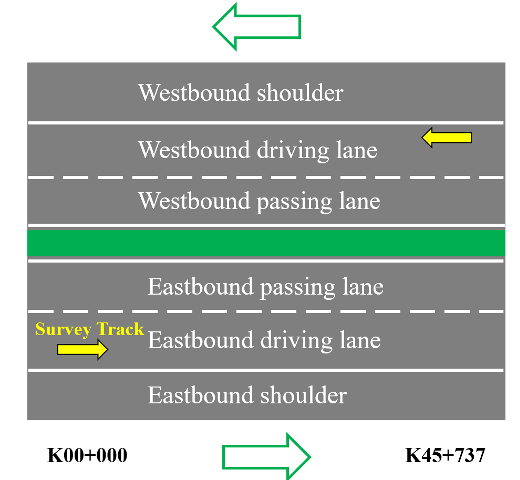}}
\label{fig_12a}
\hfil
\subfloat[Raw data formats]{\includegraphics[width=4.5in]{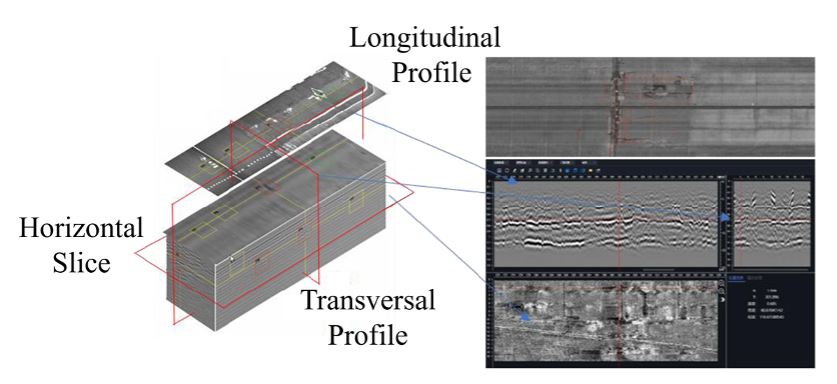}
\label{fig_12b}}
\caption{Diagram of GPR surveys on the pavement sections.}
\label{fig_12}
\end{figure*}

\begin{figure*}[!t]
\centering
\subfloat[Layer thickness=10 cm]{\includegraphics[width=3.4in]{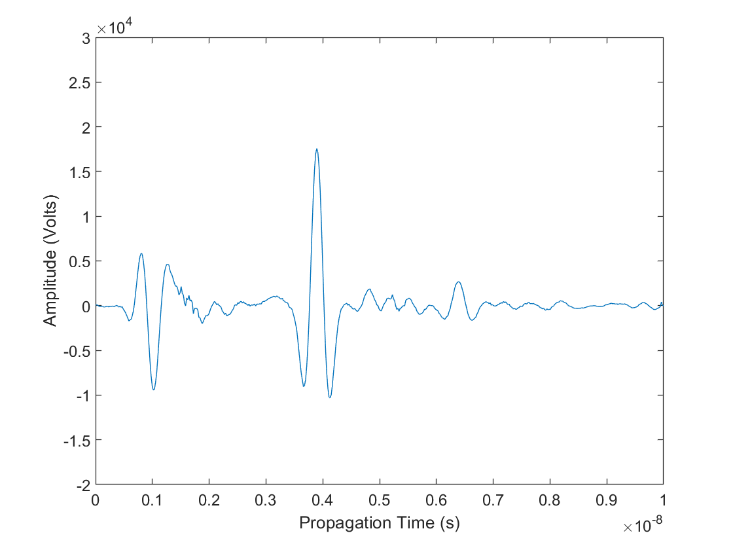}}
\label{fig_14a}
\hfil
\subfloat[Layer thickness=15 cm]{\includegraphics[width=3.4in]{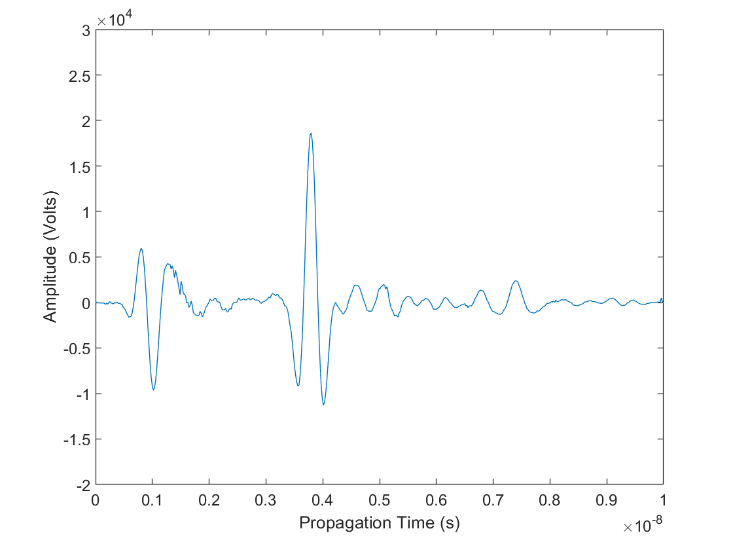}
\label{fig_14b}}
\caption{Signal waveforms using the surface reflection method.}
\label{fig_14}
\end{figure*}

\begin{figure*}[!t]
\centering
\subfloat[Layer thickness=10 cm]{\includegraphics[width=3.4in]{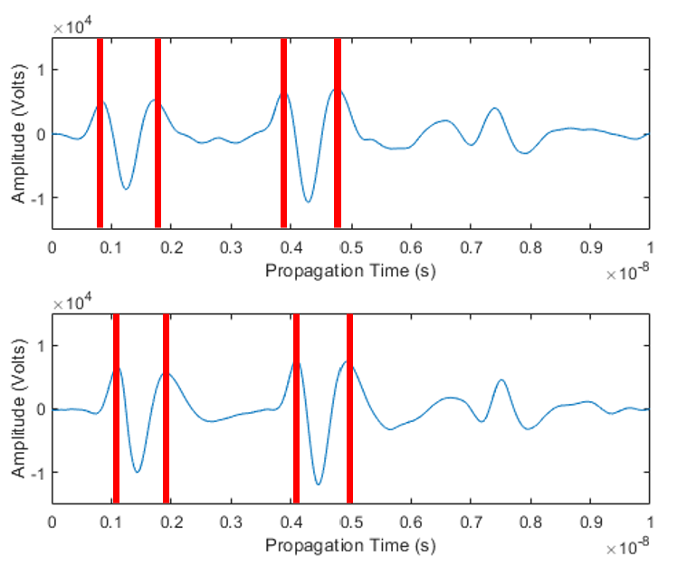}}
\label{fig_15a}
\hfil
\subfloat[Layer thickness=15 cm]{\includegraphics[width=3.4in]{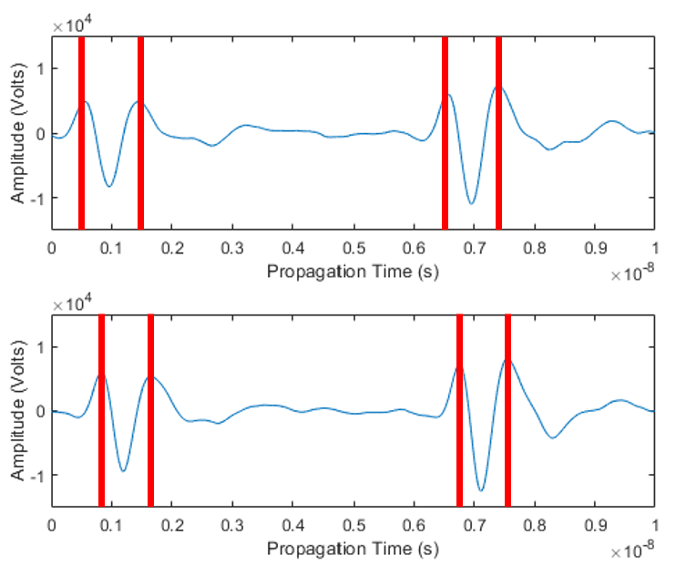}
\label{fig_15b}}
\caption{Signal waveforms using the XCMP method.}
\label{fig_15}
\end{figure*}

In-situ cores were drilled at several locations of the survey pavement sections. The layer thicknesses from cores were measured from a certified laboratory. GPR signals using surface reflection and XCMP methods were obtained at core locations, as shown in Figures \ref{fig_14} and \ref{fig_15}. Only the surface reflection amplitudes from Figure \ref{fig_14} were used to calculate the dielectric constant values using the surface reflection method in Equations \ref{eqn_1}-\ref{eqn_3}. In Figure \ref{fig_13}, the TOF between the surface and bottom reflections at the layer interface was used to calculate the thickness using the XCMP method. Note that sublayer interface reflections are invisible in the XCMP signals due to the small dielectric contrasts of sublayers. Hence, only the entire asphalt layer thickness was predicted. The proposed edge detection method was utilized to locate the reflections and predict the TOFs automatically. The optimal threshold was identified using the value from the simulation study. This is valid because the normalization process during the edge detection eliminates the effect of signal gain, which is the main difference between the signals from the simulation study and field test data. The detected reflection boundaries are highlighted in Figure \ref{fig_15}. The bulk dielectric constant and thickness were predicted using the XCMP method.

\begin{figure*}[!t]
\centering
\subfloat[Thickness values]{\includegraphics[width=3.4in]{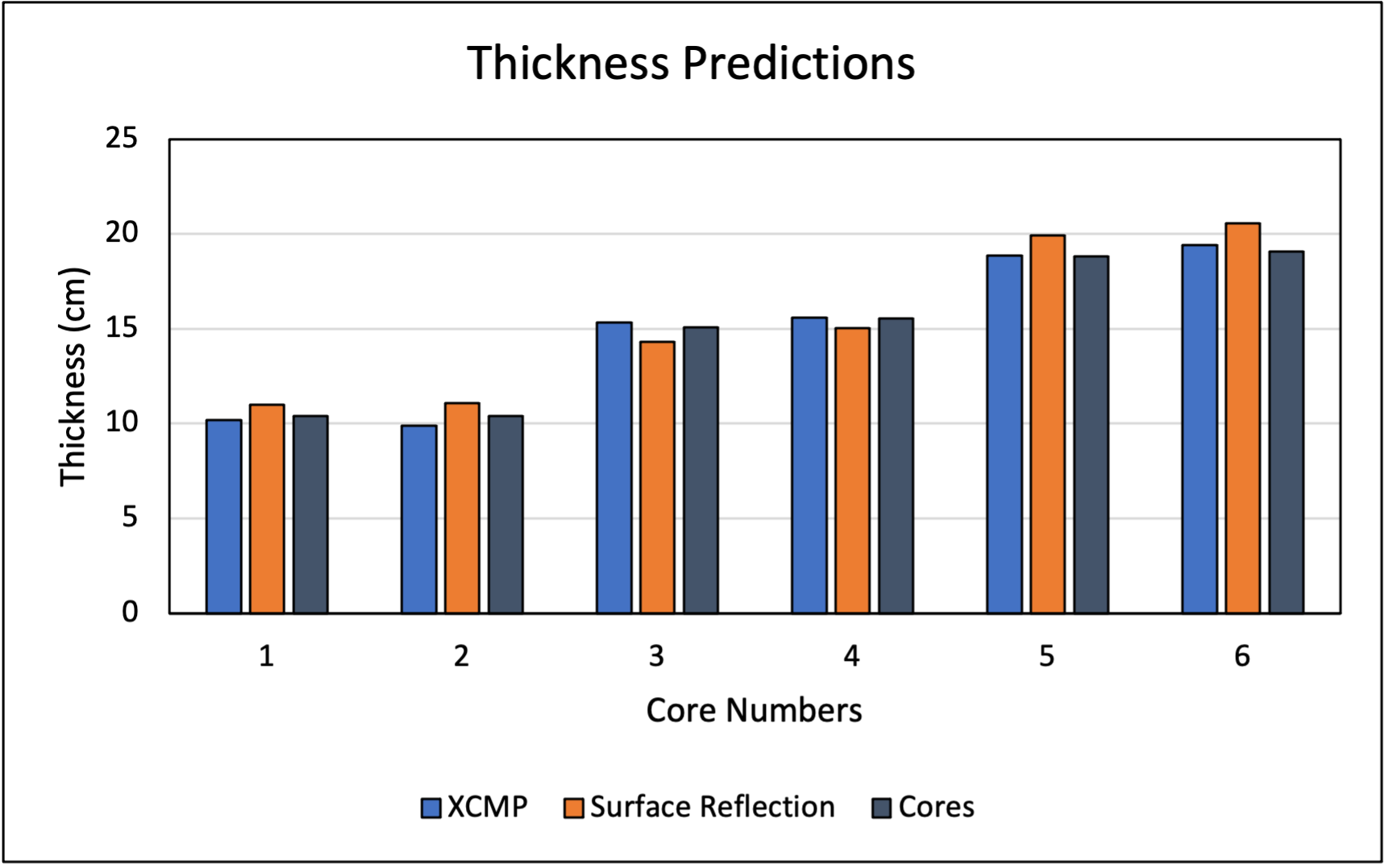}}
\label{fig_16a}
\hfil
\subfloat[Prediction errors by comparing GPR results to core measurements]{\includegraphics[width=3.4in]{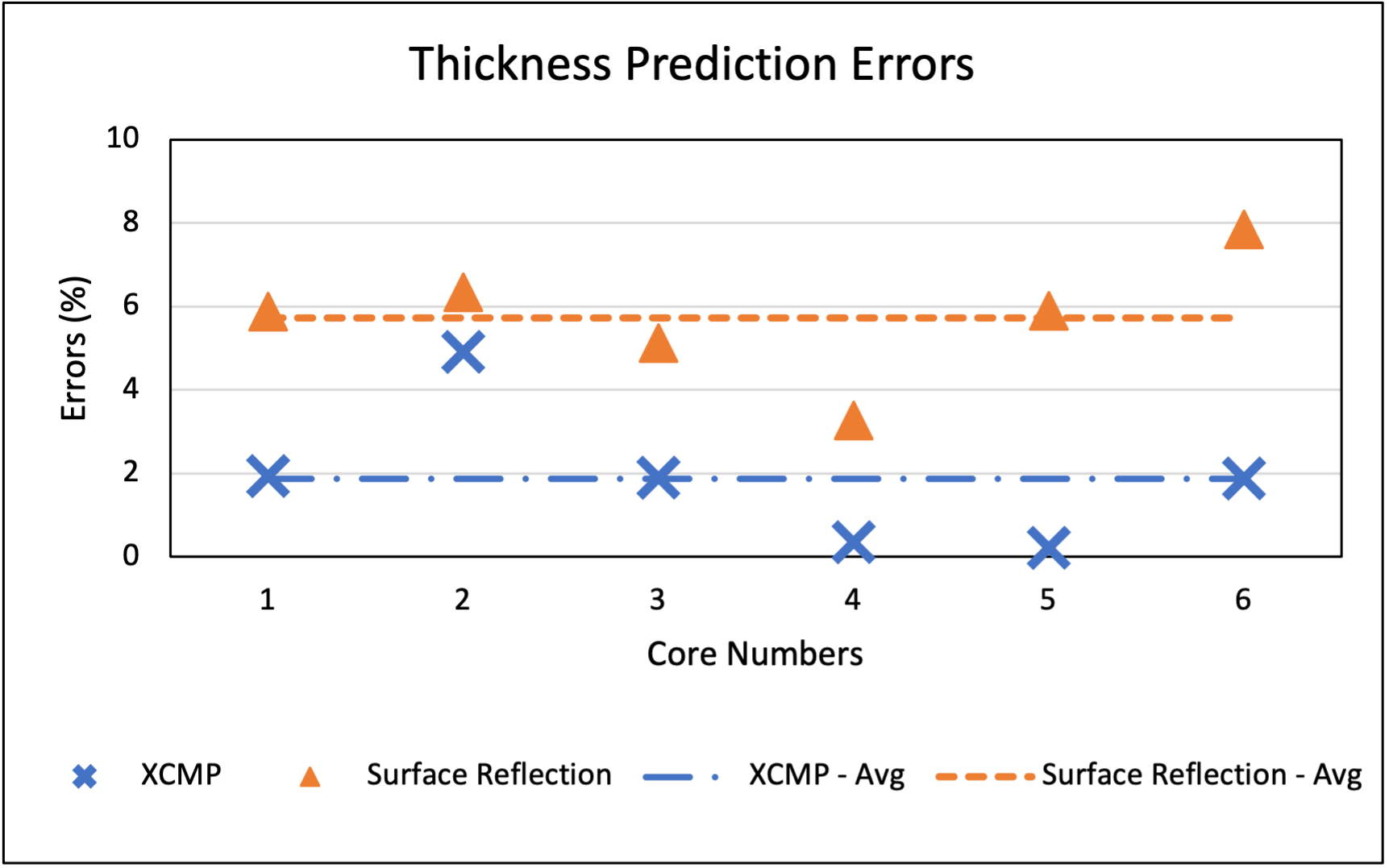}
\label{fig_16b}}
\caption{Thickness predictions from GPR and in-situ cores.}
\label{fig_16}
\end{figure*}

The predicted thickness results using GPR based on the surface reflection and XCMP methods, as well as the core measurements, are summarized in Figure \ref{fig_16}\emph{(a)}. Thickness prediction errors by comparing GPR measurements to those from cores are shown in Figure \ref{fig_16}\emph{(b)}. When layer thickness is less than 15 cm, the surface reflection method is comparable to the XCMP method in prediction accuracy, for example, at Core 2. This is caused by insufficient time resolution to capture the accurate TOFs for both antenna pairs due to the limited layer thickness. The thickness prediction error using the XCMP method is significantly lower than the surface reflection method at Cores 4, 5, and 6 where the thickness is large. This is because of the potential degradation in the thick asphalt pavement layer during the in-service period, leading to dielectric property variation through the depth. The average prediction errors are 1.86\% and 5.73\% compared to in-situ core measurements using the XCMP and surface reflection methods, respectively. The fluctuations of thickness prediction accuracies using the XCMP may come from the limited A-Scan signal sampling rate, which is 512 sampled by 10 ns. This could be addressed by increasing the A-Scan sampling rate to obtain a higher TOF estimation accuracy.

In summary, compared to the conventional surface reflection method, the XCMP method is recommended for thickness measurements of in-service pavements containing multi-lifts or pavements with non-uniform properties through depth. For the XCMP method conducted in this study, the sampling rate was 1024 samples/scan. Increasing the sampling rate using more advanced GPR hardware may lead to higher TOF estimation accuracy and more accurate thickness prediction results. Using a monostatic air-coupled or bi-static system with an adjustable separation distance between transmitter and receiver may increase the difference between $\Delta t_1$ and $\Delta t_2$, improving the thickness prediction accuracy. The proposed method could simultaneously predict several layer dielectric constants. Separating two layers and predicting thickness separately require clear reflections at interfaces between two layers, which can be achieved when layer dielectric constant contrast is large. This requirement is similar to that of the reflection method. 

\section{Conclusions}
This study investigates the factors affecting the prediction accuracy of the dielectric constant  using the XCMP method. A modified XCMP method is proposed to allow real-time thickness prediction of in-service asphalt pavement with dielectric property variation through the depth. A sensitivity analysis was performed to study the effects of layer dielectric constant variation, geometric information, and TOFs on the thickness prediction accuracy using the XCMP method, accomplished based on numerical simulations using gprMax. A modified XCMP method was proposed using the edge detector to locate the asphalt layer interface reflection signals for TOF estimations. This could allow real-time estimations of asphalt layer dielectric constant and thickness. The edge detection accuracy (EDA) was evaluated to identify an optimal sensitivity threshold to execute the edge detector automatically. The proposed method was validated via field tests using a multi-channel GPR system. The thickness predictions using the proposed method were compared to in-situ core measurements. The conclusions are summarized below.
\begin{enumerate}
    \item Dielectric constant and thickness predictions using the XCMP method are more accurate than the conventional surface reflection method with asphalt layer dielectric property variation through the depth. For the XCMP method applications, the estimation accuracy of TOFs from antenna pairs has a significantly higher effect on the prediction accuracy than the geometric information (antenna offset and geometric center height).
    \item Accurate estimation of TOFs is necessary when using the XCMP method. A modified method based on edge detection can be utilized in real time to estimate TOF. The EDA result shows the optimal sensitivity threshold as 0.01 to allow the automatic operation of the proposed edge detector. This threshold is independent of the signal gain due to the normalization process. At least two air-coupled antennas or a multi-channel antenna array with the same frequency bandwidths are required.
    \item Field test results show that the modified XCMP method is recommended for thickness measurements of in-service pavements with non-uniform properties through depth. The average prediction errors compared to in-situ core measurements are 1.86\% and 5.73\% using the XCMP and conventional surface reflection methods, respectively. The modified XCMP can be executed during data collection to provide real-time thickness prediction results.
    \item In future research, increasing the sampling rate may lead to higher TOF resolution, contributing to more accurate thickness prediction results. This can be realized by developing GPR data collection systems with high sampling rates in A-Scans. A monostatic air-coupled and bi-static system with an adjustable separation distance between the transmitter and receiver is recommended. The proposed method could simultaneously predict several layer dielectric constants. Separating two layers and predicting each thickness requires clear reflections at interfaces between two layers, which can be achieved when layer dielectric constant contrast is significant. This requirement is similar to that of the reflection method. The proposed method relies on the TOF of signal within the asphalt layer, which may be affected by internal moisture caused by rainfalls, the curing process of recycled asphalt mixtures, or water spray during the compaction process. This effect should be corrected by applying the proposed method with moisture in the asphalt pavement.
\end{enumerate}

\section*{Acknowledgment}
This research was supported by the National Key Research and Development Project (grant numbers 2020YFB1600102 and 2020YFA0714302) and the National Natural Science Foundation of China (52308444). 

\ifCLASSOPTIONcaptionsoff
  \newpage
\fi



\bibliographystyle{./IEEEtran}
\bibliography{./IEEEabrv,./IEEEexample}
\end{document}